\documentclass[twocolumn]{aastex63}
\usepackage[utf8]{inputenc}

\usepackage{amsmath}

\shorttitle{A 28~h Cessation of Accretion in YY Dra}
\shortauthors{Hill et al}

\newcommand{\orbit}{$\Omega$}
\newcommand{\spin}{$\omega$}
\newcommand{\beat}{$\omega$-$\Omega$}
\newcommand{\doubleorbit}{2$\Omega$}
\newcommand{\doublespin}{2$\omega$}
\newcommand{\doublebeat}{2($\omega$-$\Omega)$}
\newcommand{\TESS}{\textit{TESS}}

\begin{document}

\title{Hitting a New Low: The Unique 28 h Cessation of Accretion in the TESS Light Curve of YY Dra (DO Dra)}

\author[0000-0003-1866-7607]{Katherine L. Hill}
\affiliation{Department of Physics, University of Notre Dame, Notre Dame, IN 46556, USA}

\author[0000-0001-7746-5795]{Colin Littlefield}
\affiliation{Department of Physics, University of Notre Dame, Notre Dame, IN 46556, USA}
\affiliation{Department of Astronomy, University of Washington, Seattle, WA 98195, USA}
\affiliation{Bay Area Environmental Research Institute, Moffett Field, CA 94035 USA}

\author[0000-0003-4069-2817]{Peter Garnavich}
\affiliation{Department of Physics, University of Notre Dame, Notre Dame, IN 46556, USA}

\author{Simone Scaringi}
\affiliation{Centre for Extragalactic Astronomy, Department of Physics, Durham University, South Road, Durham, DH1 3LE}

\author[0000-0003-4373-7777]{Paula Szkody}
\affiliation{Department of Astronomy, University of Washington, Seattle, WA 98195, USA}

\author{Paul A. Mason}
\affiliation{New Mexico State University, MSC 3DA, Las Cruces, NM, 88003, USA}
\affiliation{Picture Rocks Observatory, 1025 S. Solano Dr. Suite D., Las Cruces, NM 88001, USA}

\author[0000-0001-6894-6044]{Mark R. Kennedy}
\affiliation{Department of Physics, University College Cork, Cork, Ireland}
\affiliation{Jodrell Bank Centre for Astrophysics, Department of Physics and Astronomy, The University of Manchester, M19 9PL, UK}

\author{Aarran W. Shaw}
\affiliation{Department of Physics, University of Nevada, Reno, NV 89557, USA}

\author{Ava E. Covington}
\affiliation{Department of Physics, University of Nevada, Reno, NV 89557, USA}

\correspondingauthor{Katherine Hill, Colin Littlefield}
\email{khill9@nd.edu, clittlef@alumni.nd.edu}

\begin{abstract}
    We present the Transiting Exoplanet Surveying Satellite (\TESS) light curve of the intermediate polar YY Draconis (YY Dra, also known as DO Dra). The power spectrum indicates that while there is stream-fed accretion for most of the observational period, there is a day-long, flat-bottomed low state at the beginning of 2020 during which the only periodic signal is ellipsoidal variation and there is no appreciable flickering. We interpret this low state to be a complete cessation of accretion, a phenomenon that has been observed only once before in an intermediate polar. Simultaneous ground-based observations of this faint state establish that when accretion is negligible, YY Dra fades to $g=17.37\pm0.12$, which we infer to be the magnitude of the combined photospheric contributions of the white dwarf and its red dwarf companion. Using survey photometry, we identify additional low states in 2018-2019 during which YY Dra repeatedly fades to---but never below---this threshold. This implies relatively frequent cessations in accretion. Spectroscopic observations during future episodes of negligible accretion can be used to directly measure the field strength of the white dwarf by Zeeman splitting. Separately, we search newly available catalogs of variable stars in an attempt to resolve the long-standing dispute over the proper identifier of this system.
    
\end{abstract}

\section{Introduction}
A cataclysmic variable (CV) is a binary star system that features mass transfer via Roche lobe overflow onto a white dwarf (WD). As the material has too much angular momentum to accrete directly onto the WD, it instead forms an accretion disk around the WD. A thorough review of CVs can be found in \cite{Warner95} and \cite{Hellier01}. 

Also known as DQ Herculis stars, intermediate polars (IPs) are a subset of CVs in which the WD has a comparatively weak magnetic field and rotates asynchronously \citep[for a review, see][]{Patterson94}. In most well-studied IPs, the magnetic field disrupts the inner accretion disk, forcing the gas to travel along magnetic field lines until it impacts the WD near its magnetic poles; these systems are commonly said to be ``disk-fed.'' In principle, the magnetic field can disrupt the disk altogether, resulting in ``diskless'' or ``stream-fed'' accretion. The simulations shown in Fig.~3 of \citet{norton04} provide an excellent visualization of stream-fed accretion flows compared to their disk-fed counterparts. However, in practice, long-term diskless accretion is rare amongst known IPs. V2400~Oph \citep{buckley95, buckley97} is the best-known example of persistently diskless IP, and Paloma \citep{schwarz, joshi} is a strong candidate as well. A few IPs show evidence of an accretion disk at some epochs while being diskless at others, with FO~Aqr being one proposed example \citep{Littlefield20}. In some IPs, the stream can overflow the disk until it impacts the magnetosphere creating a hybrid of diskless and disk-fed accretion.

The subject of this study, YY Draconis (hereafter, YY Dra),\footnote{YY Dra is also known as DO Draconis (DO Dra) due to an ambiguity in identification, as explained in detail by \citet{patterson87}. In our Appendix, we discuss new evidence that YY~Dra is the correct identifier.} is an IP with a lengthy observational history following its identification by \citet{patterson82} as the optical counterpart of the X-ray source 3A\ 1148+719. Its status as an IP was established through optical and X-ray studies by \citet{Patterson92} and \citet{patterson93}, respectively. The optical data from \citet{Patterson92} showed wavelength-dependent periods of 275~s and 265~s, each of which was identified as the second harmonic of a pair of fundamental frequencies of variation (550~s and 529~s respectively). The subsequent X-ray study by \citet{patterson93} found an X-ray counterpart to the 265~s signal, which the authors attributed to the rotation of the WD. They argued the two accretion regions contribute almost equally to the light curve, causing the fundamental WD spin frequency (\spin = 163 cycles d$^{-1}$) to have a low amplitude compared to its second harmonic. Their identification of \spin\ meant that the 550~s signal is the spin-orbit beat period (\beat = 157 cycles d$^{-1}$). The dominance of the harmonics is consistent with equatorial accretion regions that contribute nearly equally to the light curve \citep{szkody02}. 

A spectroscopic study by \cite{Matteo91} established the system's orbital period to be 3.96~h and estimated the orbital inclination to be $i = 42^{\circ}\pm5^{\circ}$. \citet{haswell} and \citet{joshi_yy_dra} refined the orbital period, and the \citet{joshi_yy_dra} measurement of 0.16537424(2)~d (\orbit = 6.061 cycles d$^{-1}$) is the most precise value in the previous literature. 
The Gaia Early Data Release 3 (EDR3) distance of YY Dra is 196.6 $\pm$ 1.1 pc \citep{BJ21}.

\section{Observations}

\subsection{\TESS}
The Transiting Exoplanet Survey Satellite (\TESS) observed YY Dra from 2019 December 25 until 2020 February 28 in Sectors 20 and 21. During these 65 days, it observed the system at a two-minute cadence across a broad bandpass ($\sim$600-1,000~nm) that covers the red and near-infrared spectrum, including the $R$ and $I$ bands. For a deeper description of \TESS, see \citet{Ricker2015}. Using the Python package {\tt lightkurve} \citep{Lightkurve20}, we downloaded the \TESS\ light curve, which has nearly continuous coverage with the exception of three $\sim$2-day gaps due to data downlinks. 

The flux measurements from the \TESS\ pipeline come in two forms: simple-aperture photometry (SAP) and pre-conditioned simple-aperture photometry (PDCSAP). The latter is a processed version of the former that attempts to remove systematic trends and the effects of blending. Although the elimination of these problems is desirable in principle, \citet{littlefield21} argued that the PDCSAP light curve of the IP TX~Col removed an outburst that was present in the SAP light curve. With this issue in mind, we plotted the \TESS\ PDCSAP and SAP fluxes as a function of the simultaneous Zwicky Transient Facility (ZTF) \citep{Smith19} $g$ flux, repeating the procedure separately for each sector. Our objective was to determine which version of the light curve better correlated with the contemporaneous ground-based data. 

Following the exclusion of a single discrepant ZTF measurement, we determined that the SAP flux was much more strongly correlated with the ZTF flux than was the PDCSAP flux. The difference was especially pronounced during Sector 21, when the best-fit linear regressions had coefficients of determination of $r^{2}_{SAP} = 0.92$ and $r^{2}_{PDC} = 0.24$ for the SAP and PDCSAP light curves, respectively. These statistics indicate that 92\% of the variation in the SAP light curve but only 24\% of the variation in the PDCSAP data can be explained by the variations in the ZTF flux. During Sector 20, we found that $r^{2}_{SAP} = 0.26$ and $r^{2}_{PDC} = 0.02$. The higher values of $r^2$ during Sector 21 are probably because the range in brightness was much greater than in Sector 20. A visual inspection of the residuals from the linear fits confirmed that the low value of $r^{2}_{PDC}$ in both sectors was attributable to a weak correlation as opposed to a strong, but non-linear, relationship. Consequently, we elected to use the SAP light curve in our analysis. 

\begin{figure*}
\centering
\includegraphics[width=\textwidth]{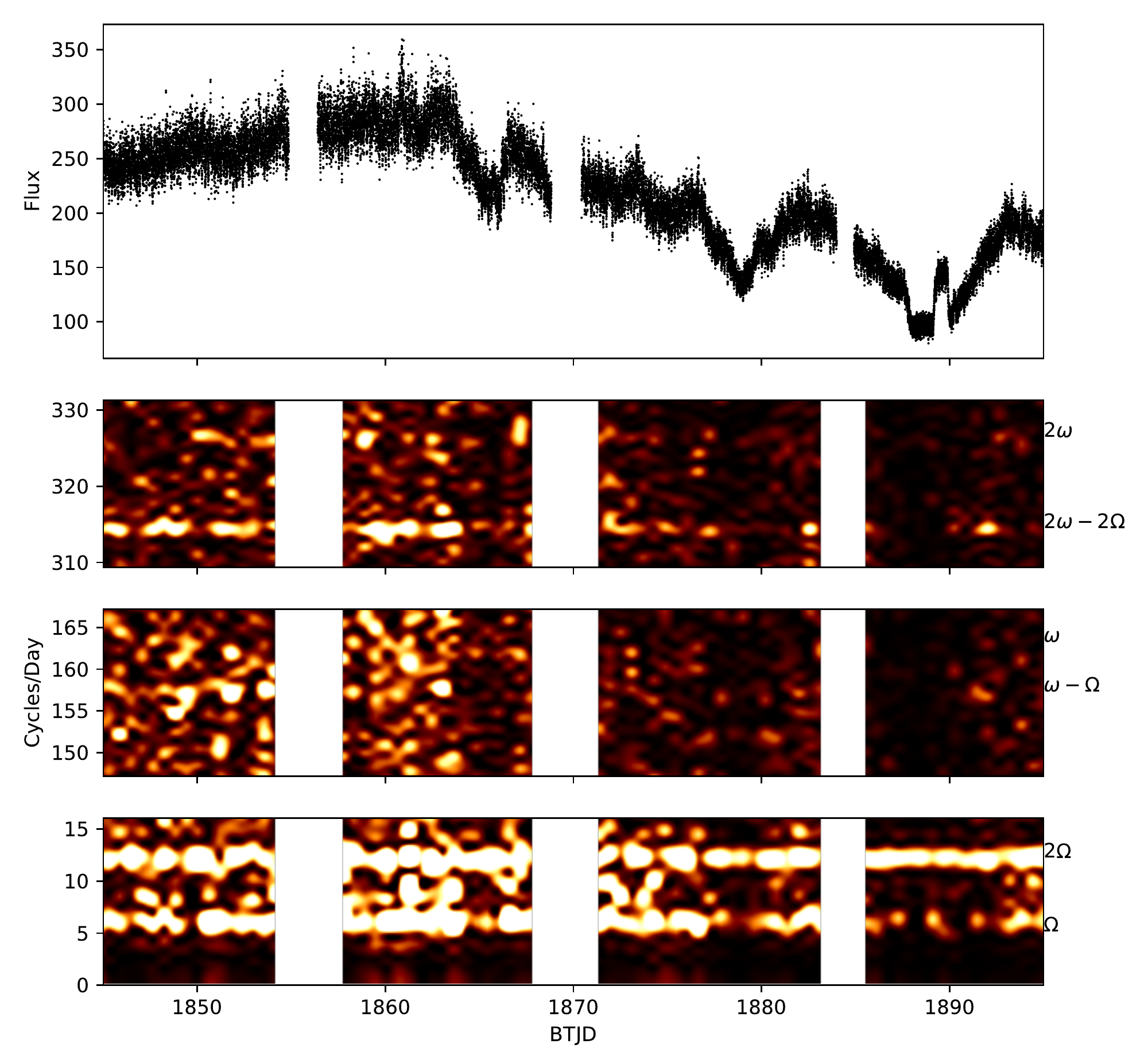}
\caption{{\bf Top Panel:} \TESS\ light curve of YY Dra, showing the SAP flux from the mission's pipeline. The deep low state described in the text is apparent at BTJD=1888. {\bf Bottom Three Panels:} Slices of YY Dra's two-dimensional power spectrum near frequencies of interest, following the subtraction of a smoothed version of the light curve to suppress red noise. All periodic variation ceases during the deep low state except for the \doubleorbit\ signal, which we attribute to ellipsoidal variations by the donor star. The same intensity color map is used for all three panels.}
\label{fig:3DLC}
\end{figure*}

\begin{figure*}
\centering
\includegraphics[width=\textwidth]{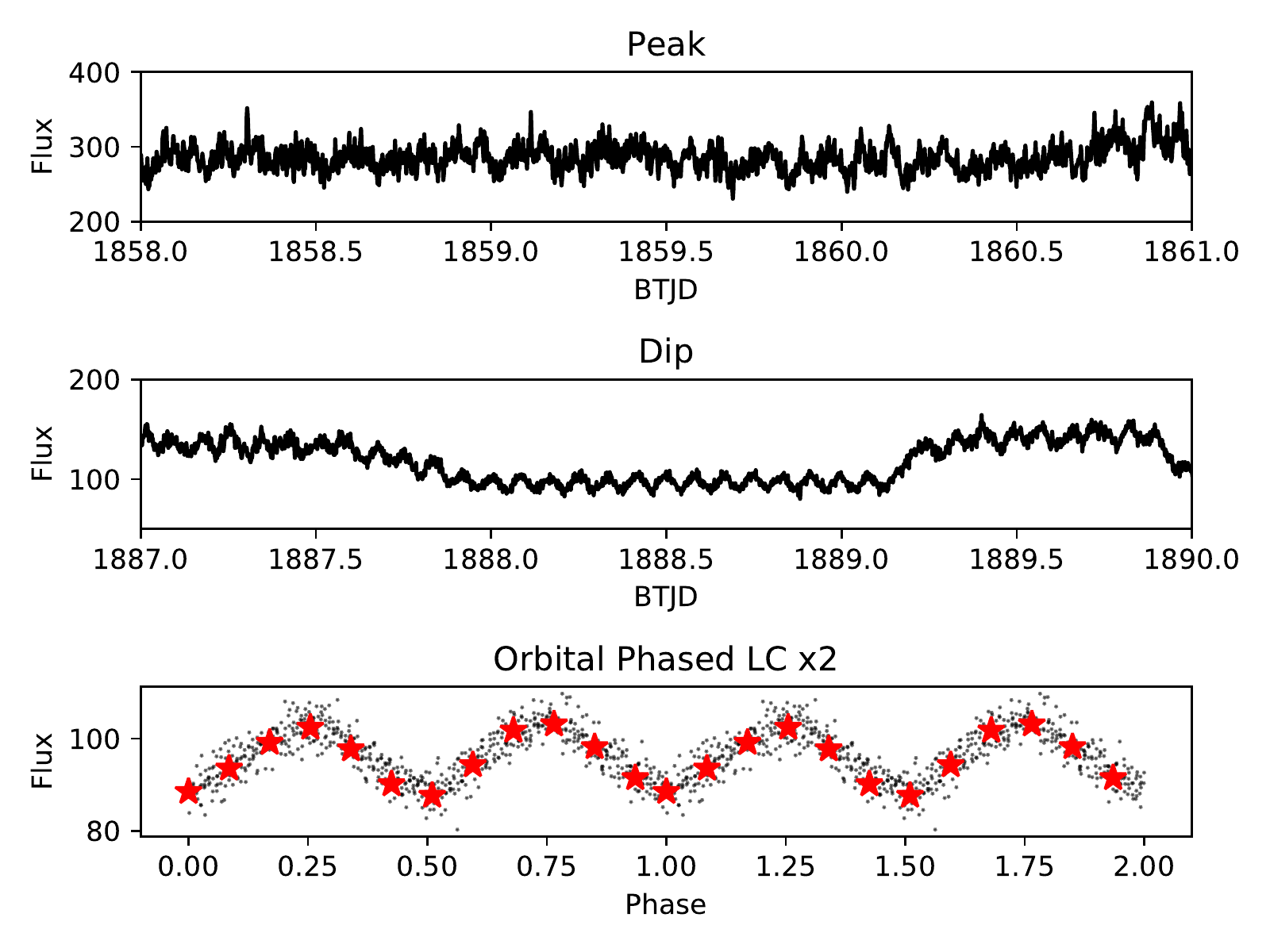}
\caption{Zoomed section of both a high and low state in YY Dra's \TESS\ light curve. {\bf Top:} Two-day portion in the \TESS\ light curve during the slow brightening at the beginning of the observation. This is representative of the first three weeks of the \TESS\ light curve. {\bf Middle:} 1.2~d deep low state during which accretion ceased. During this low state, the only periodic variability is from ellipsoidal variations. {\bf Bottom:} YY Dra's orbital light curve during the deep low state between BTJD = 1888-1889, phased according to Eq.~\ref{ephemeris}. The red stars represent bins with a width of 0.085. The data are repeated for clarity.} 
\label{fig:High.Low.PhasedLC}
\end{figure*}

The top panel of Figure~\ref{fig:3DLC} shows the full \TESS\ light curve, while Figure~\ref{fig:High.Low.PhasedLC} shows selected segments to give a sense of the data quality. Throughout this paper, including Figure~\ref{fig:3DLC}, we use the Barycentric \TESS\ Julian Date (BTJD) convention to express time, which is defined as BTJD = BJD $-$ 2457000, where BJD is the Barycentric Julian Date in Barycentric Dynamical Time.

\subsection{Survey Photometry}

\begin{figure*}
\centering
\includegraphics[width=\textwidth]{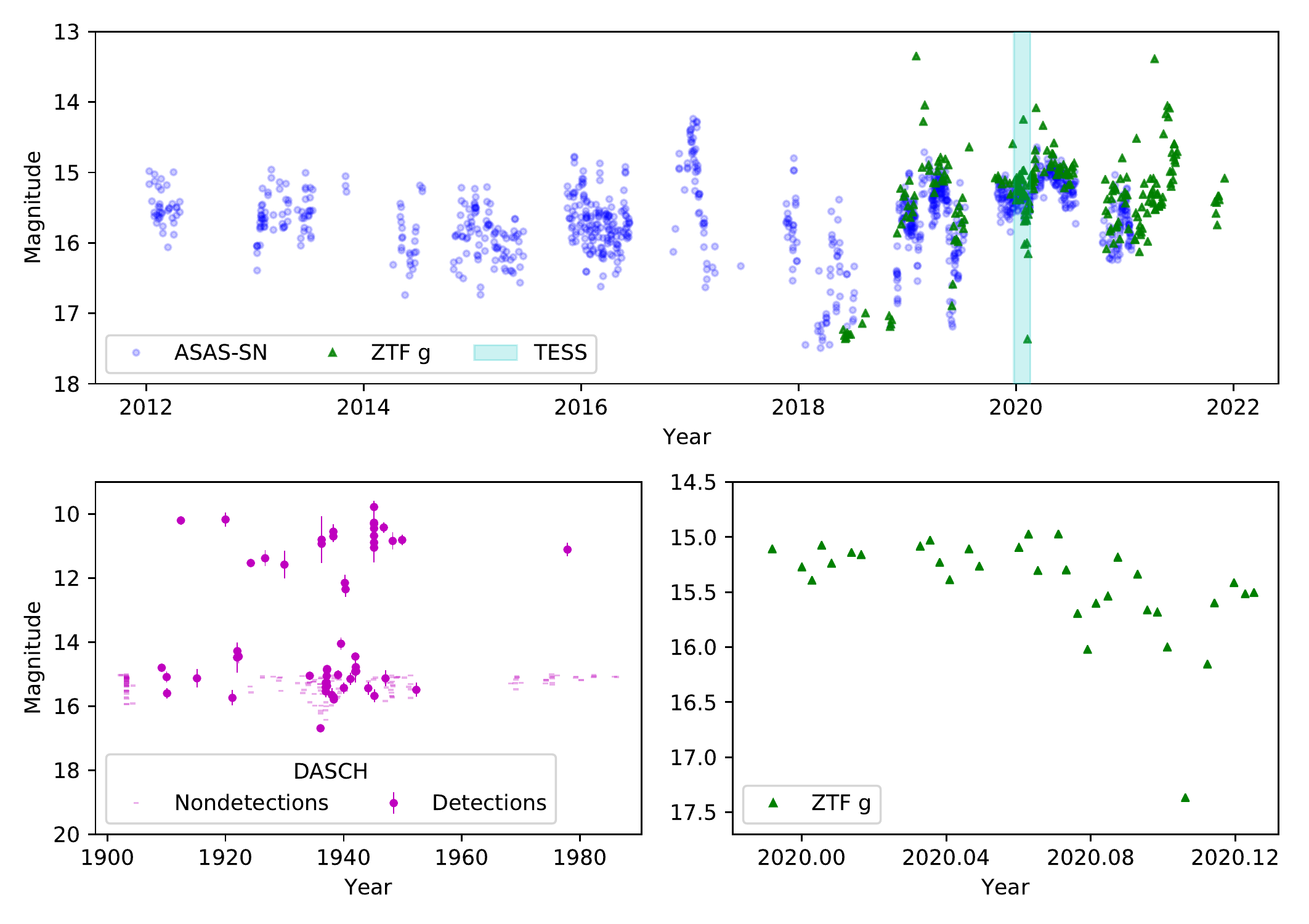}
\caption{YY Dra light curves from ASAS-SN, ZTF, and DASCH.  {\bf Top:} Overlaid ASAS-SN (blue circles) and ZTF (green triangles) light curves with including a deep low state observed by \TESS\ shown in the cyan shaded region. The ASAS-SN light curve is relatively uniform for the first few years. However, multiple intermittent low states begin to appear around 2018-2019. The observed low states reach a consistent minimum magnitude. {\bf Bottom-left:} The DASCH light curve, with detections shown as circles and nondetections as dashes. As explained in the text, nondetections with limits brighter than magnitude 15 have been excluded. The quiescent magnitude is comparable to that observed by ZTF. While there are numerous outbursts throughout the observational period, no low states comparable to those in the ASAS-SN and ZTF data are present.  {\bf Bottom-right:} Enlargement of the ZTF observations obtained during the \TESS\ light curve. The faintest datum ($g=17.37\pm0.12$) was obtained during the deep low state, when accretion was negligible.
}
\label{fig:DiffLC}
\end{figure*}

We compared the \TESS\ light curve to a long-term light curve (Figure~\ref{fig:DiffLC}) assembled from the Harvard Digital Access to a Sky-Century at Harvard project \citep[DASCH;][]{Grindlay09}, the All-Sky Automated Survey for Supernovae \citep[ASAS-SN; ][]{Shappee14,Kochanek17}, and the ZTF. The DASCH light curve used the default quality flags of the project's pipeline, and its magnitudes are appoximately equal to the Johnson $B$ band. We used only $g$ magnitudes from ASAS-SN and ZTF.


\section{Analysis}

\begin{figure*}
\centering
\includegraphics[width=\textwidth]{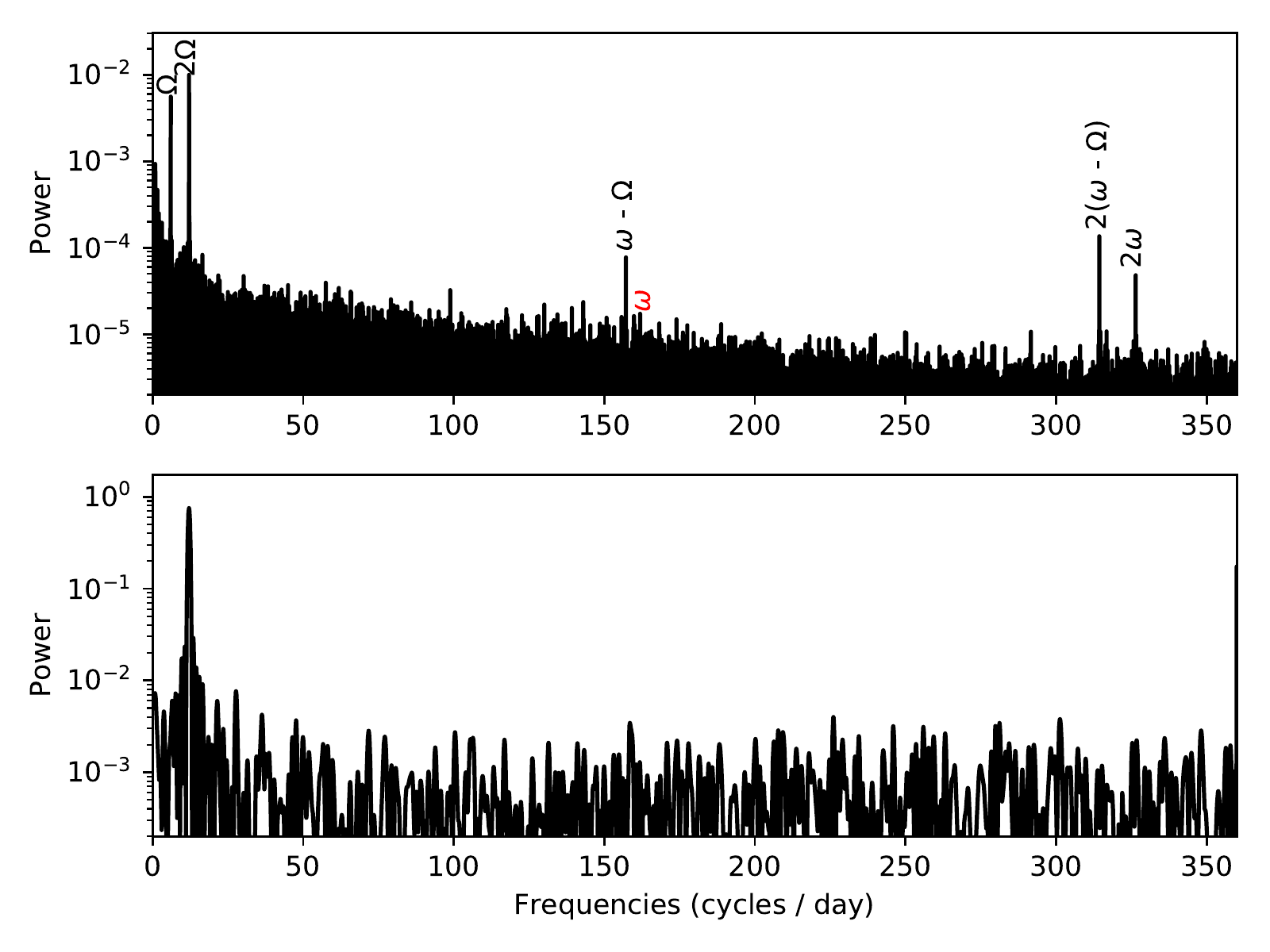}
\caption{ {\bf Top:} YY Dra's power spectrum of the entire \TESS\ light curve, with very clear \orbit, \doubleorbit, \beat, \doublebeat, and \doublespin\ frequencies. The spin frequency \spin\ was below the noise level, but its expected position is labeled in red. The comparatively high intensity of \beat\ and its harmonics relative to \spin\ and \doublespin\ is characteristic of stream-fed accretion in the \citet{Ferrario99} models. {\bf Bottom:} Power spectrum during the 28~h low state shown in Fig.~\ref{fig:High.Low.PhasedLC}. The only remaining periodic variation occurs at \doubleorbit\ and is attributable to ellipsoidal variations. 
}
\label{fig:PwrSpectrum}
\end{figure*}

\subsection{\TESS\ light curve}

During the \TESS\ observation, YY Dra was comparatively bright for the first 20 days ($g\sim15.2$; Fig.~\ref{fig:DiffLC}), showing a slow, rising trend in the \TESS\ data. However, for the remainder of the light curve, it generally exhibited a gradual fade, punctuated by three obvious dips (Figure~\ref{fig:3DLC}, top~panel). The slow fade began near BTJD=1865, leading into the first of the three dips at BTJD = 1868. Although YY~Dra quickly recovered from this dip, the overall fading trend continued, and another, deeper dip took place at BTJD = 1878. The second dip consisted of a relatively rapid fade over $\sim2.5$~d and thereafter recovered within 3.5~d, lasting about a week in total. 

The third dip is the most remarkable feature of the light curve. Unlike the first two dips, its deepest portion is flat-bottomed, lasts for just 1.2~d, and was observed by ZTF to be $g=17.37\pm0.12$, approximately 2~mag fainter than the beginning of the \TESS\ light curve. The middle panel of Figure~\ref{fig:High.Low.PhasedLC} shows a 3~d section of the light curve centered on this dip. The deepest part of the low state shows smooth, sinusoidal variation, the nature of which which we explore in Sec.~\ref{sec:2Dpower}. This low state began with a rapid 33\% drop in brightness lasting eight hours at BTJD = 1888, and it ended with a burst that lasted for 16~h. After a rapid fade from the burst, YY~Dra began a slow recovery, but it never completely recovered to its starting brightness during the \TESS\ observation.

In the following subsections, we use power spectral analysis to gain insight into these behaviors. 

\subsection{1D Power Spectrum}
\label{sec:1Dpower}

We computed a Lomb-Scargle power spectrum \citep{LOMB,SCARGLE} to examine YY~Dra's periodic variations between 0.67 cycles~d$^{-1}$ and 360 cycles~d$^{-1}$. The \orbit, \doubleorbit, \doublebeat, and \doublespin\ frequencies from \citet{haswell} are easily identified in this power spectrum (Figure~\ref{fig:PwrSpectrum}). Because of the relatively short \TESS\ baseline, the measured periods from the power spectrum are statistically indistinguishable from those in \citet{haswell}.

The power spectrum of an IP contains important information about the system's structure  \citep{warner86, Ferrario99, murray}. For example, \citet{Ferrario99} computed theoretical optical power spectra for stream-fed and disk-fed IPs. They predicted that the spin period should tend to be very apparent during disk-fed accretion because the accretion disk encircles the entire white dwarf, providing each magnetic pole with a uniform reservoir of matter from which to accrete, regardless of the WD's rotational phase. In stream-fed accretion, however, the white dwarf accretes from a fixed region in the binary rest frame: a single, focused point where the stream from the donor star encounters the magnetic field. In sharp contrast to disk-fed accretion, the accretion rate onto any one pole will vary with the WD's rotation. 

In YY Dra, stream-fed accretion could explain why there is a very conspicuous \beat\ and \doublebeat\ and a rather inconspicuous \spin. The \citet{Ferrario99} models predict that stream-fed accretion will tend to shift power into \beat\ and its harmonics, and in YY~Dra, the amplitudes of \beat\ and \doublebeat\ are much higher than \spin\ and \doublespin. This distribution of power is not accounted for in the \citet{Ferrario99} disk-fed models, but it is consistent with their stream-fed models, particularly at lower inclinations. Furthermore, the abrupt disappearance of \orbit\ during a 28~h cessation of accretion during the \TESS\ observation (Sec.~\ref{sec:2Dpower}) reveals that this frequency is produced by accretion (as opposed to the changing aspect of the donor star), which is consistent with a prediction by \citet{Ferrario99} that stream-fed accretion will often shift power into \orbit.



Although we believe that the \citet{Ferrario99} stream-fed accretion models offer the best explanation of the observed power spectrum, we also considered two other scenarios to account for YY~Dra's comparatively strong \beat\ and \doublebeat\ signals. The first was proposed by \citet{warner86}, who argued that stationary structures in the binary frame can reprocess X-rays into the optical at a frequency of \beat; we presume that this effect could also produce a signal at \doublebeat\ if both poles participate in this process. However, one challenge with applying this scenario to YY~Dra is that at its inclination of $i=42^{\circ}$ \citep{Matteo91}, we would expect at least a modest amplitude modulation of the reprocessed optical radiation across the binary orbit, shifting power into the upper and lower orbital sidebands of both \beat\ and \doublebeat \citep{warner86}. This predicted distribution of power is inconsistent with the observed power spectrum in Fig.~\ref{fig:PwrSpectrum} and suggests that the region in which the optical \beat\ and \doublebeat\ signals originate is uniformly visible across the orbit. While not absolutely conclusive, this reasoning disfavors reprocessed X-rays as the origin of \beat\ and \doublebeat\ in the \TESS\ data.

Another alternative explanation that we considered involves the possibility of spiral structure within the accretion disk. \citet{murray} computed theoretical power spectra for IPs in which spiral arms in the disk extend far enough inward to reach the disk's inner rim. Similar to stream-fed accretion, this scenario would break the azimuthal symmetry of the accretion flow at the magnetospheric boundary, producing power at \beat\ and/or \doublebeat. However, we disfavor this interpretation for two reasons. First, it seems unlikely that if a disk were present, the power at \beat\ and \doublebeat\ could dwarf that at \spin\ and \doublespin. Second, we would not expect spiral disk structure to be present during quiescence in a CV with YY~Dra's 4~h orbital period. While spiral structure has been detected in IP~Peg ($P_{orb}$ = 3.8~h) during its outbursts \citep{steeghs97}, the physical conditions in the disk during outburst---particularly its viscosity---are very different than in quiescence, and YY~Dra's faintness during the \TESS\ observation rules out the possibility that the system possessed an outbursting disk. 

We therefore conclude that YY Dra was predominantly stream-fed during these observations.

\subsection{2D Power Spectrum}
\label{sec:2Dpower}

Figure~\ref{fig:3DLC} presents the evolution of the power spectrum over time and offers insight into the stability of YY~Dra's periodic variations as well as their dependence on the brightness of YY Dra. Throughout the observation, \orbit\ and \doubleorbit\ were consistently the largest-amplitude signals in the two-dimensional power spectrum. Of the frequencies related to the WD's rotation, \doublebeat\ consistently had the highest amplitude, while \beat, \spin, and \doublespin\ were not reliably detected above the noise in 0.5~day segments of the light curve. The amplitude of \doublebeat\ declines in the second half of the light curve, coinciding with YY~Dra's gradual fade.

During the 28~h long, flat-bottomed low state at BTJD=1888, no periodic variability is present, except for \doubleorbit, and even the flickering appears to stop. We attribute this behavior to a cessation of accretion during which only ellipsoidal variations from the companion star are present in the light curve, as seen in other CVs during zero-accretion states \citep[e.g. KR~Aur; ][]{kr_aur}. The absence of accretion during this 28~h low state distinguishes it from the low state detected by \citet{Breus17}, during which YY~Dra showed evidence of both ellipsoidal variations and additional variability at the orbital frequency, resulting in a double-humped orbital profile with unequal maxima; had accretion ceased altogether during the \citet{Breus17} observations, we would expect to observe the maxima to have been equal. While episodes of negligible accretion have been identified in other magnetic CVs, our review of the literature suggests that states of insignificant accretion are extremely rare in IPs, as we discuss in Sec.~\ref{sec:comparison}.

Lastly, we unsuccessfully searched the two-dimensional power spectrum for evidence of transient periodic oscillations (TPOs), a phenomenon identified in YY~Dra by \citet{andronov08}. The TPOs in that study consisted of a mysterious signal near 86~cycles~d$^{-1}$ whose frequency gradually decreased to 40~cycles~d$^{-1}$ over the course of three nights. Because this phenomenon in the \citet{andronov08} dataset occurred at an elevated accretion rate ($\sim1$~mag brighter than during the \TESS\ observation), the lack of TPOs in the \TESS\ data suggests that they are sensitive to the accretion rate.

\subsection{Long-Term Light Curve}

The DASCH light curve spans from 1902 to 1986 and is by far the most sparsely sampled of the three datasets in Figure~\ref{fig:DiffLC}. However, it offers a unique look at YY~Dra in the decades prior to its discovery. The DASCH data show no low states as deep as the one observed by both \TESS\ and ZTF, although one observation (at magnitude $B=16.7$ in 1936) comes fairly close. The lack of deep low states could be attributable in part to the relatively shallow limiting magnitude of many of the photographic plates. Indeed, Figure~\ref{fig:DiffLC} excludes 6,165 non-detections with limiting magnitudes brighter than 15, a threshold selected because it is too shallow to reach YY Dra's quiescent brightness. Even the non-detections with comparatively deep limits cannot meaningfully distinguish between normal quiescent variability and a low state. Additionally, there are no observations during the early 1950s and late 1960s, so any low states during that interval would have gone undetected. Finally, while low states are elusive in the century-long light curve of YY~Dra, the DASCH data do emphasize the bright, $>5$~mag outbursts of YY~Dra, a distinguishing property noted by previous studies \citep[e.g.,][]{szkody02}.

While the DASCH data provide a partial glimpse of YY~Dra's behavior across the 20th century, the ASAS-SN and ZTF survey photometry provide significantly denser observational coverage during the past decade. The ASAS-SN and ZTF $g$ light curves, both of which overlap the \TESS\ observations, are in excellent agreement with each another (Fig.~\ref{fig:DiffLC}, top panel). Although ASAS-SN has a longer baseline by six years, the ZTF data fill in some of the gaps in the ASAS-SN coverage. In 2018, YY~Dra was regularly in a low state, dropping below $g\sim17$, and while the exact duration of the low states is not clear, some of them could have lasted for months.

The ZTF data are especially critical because they include an observation of YY~Dra during its deep low state (Figure~\ref{fig:DiffLC}). This single datum therefore establishes the magnitude of YY~Dra when its accretion rate was negligible: $g = 17.37\pm0.12$. In the absence of any accretion, the optical luminosity of YY~Dra should be composed of the stellar photospheres of the WD and the companion star. As such, the ZTF measurement should represent the minimum magnitude which YY Dra can appear, at least until the WD has cooled significantly. Inspecting Figure~\ref{fig:DiffLC}, it is evident that  YY~Dra spent much of 2018 in or near such a state.\footnote{\citet{covington} examined archival observations of YY~Dra from 2018 and identified a 3~mag fade in Swift ultraviolet UVW- and UVM-band observations that coincided with a non-detection of YY~Dra in X-ray observations. Likewise, \citet{shaw2020} found no X-ray emission in a 55.4~ks NuSTAR observation of YY~Dra in July 2018. These findings provide independent support for a cessation of accretion onto the WD in 2018. }



\subsection{Updated Orbital Period}

The minima of the ellipsoidal variations in Figure~\ref{fig:High.Low.PhasedLC} are almost perfectly equal, making it difficult to identify which one corresponds to inferior conjunction of the donor star. However, the data can be reliably phased to the binary orbit using the \citet{joshi_yy_dra} orbital period and the \citet{haswell} epoch of inferior conjunction.\footnote{ We converted the epoch of the \citet{haswell} to Barycentric Julian Date in Barycentric Dynamical Time using routines in {\tt astropy}.} Because of the excellent precision of the \citet{joshi_yy_dra} orbital period (0.16537424(2)~d), this trial ephemeris has an accumulated phase error of only $\pm0.009$ at the time of the low state near BTJD=1888, so when the data are phased to it, the photometric minimum corresponding to inferior conjunction should be observed within $\pm0.009$~orbital cycles of phase 0.0. 

The phased data agree with this prediction. When phased to the aforementioned trial ephemeris, the closest minimum to phase 0.0 occurred at a nominal orbital phase of $\sim-0.02$. While this suggests that the \citet{joshi_yy_dra} uncertainty might have been slightly underestimated, the phasing of the deep-low-state light curve strongly supports our inference that ellipsoidal variations are responsible for the observed signal at \doubleorbit.

Refining the orbital period slightly to 0.16537420(2)~d corrects the phase shift from the \citet{joshi_yy_dra} orbital period. Adopting this revised orbital period, we offer an orbital ephemeris of \begin{equation}
    T_{conj}[BJD] = 2446863.4383(20) + 0.16537420(2)\times\ E,
    \label{ephemeris}
\end{equation}
where $E$ is the integer cycle count.

\section{Discussion}

\subsection{Comparison to Low States in Other IPs} \label{sec:comparison}

\citet{Garnavich88} searched for low states in three IPs in the Harvard plate collection: V1223 Sgr, FO Aqr, and AO Psc. They found that V1223~Sgr experienced a nearly decade-long low state whose depth was typically $\sim$1~mag but briefly exceeded 4~mag.
Additionally, their analysis of AO~Psc revealed a single $\sim$1~mag low state, lasting for at most several weeks. However, considering the difference in these time scales compared to YY~Dra's 28~h low state as well as the absence of contemporaneous time series photometry for the low states in the Harvard plates, it is difficult to compare the \citet{Garnavich88} low states against the low state in the \TESS\ light curve.

More recently, low-accretion states have been detected in several confirmed or suspected IPs, including V1323~Her \citep{V1323Her}, FO Aqr \citep{Littlefield16, Kennedy17}, DW Cnc \citep{Montero20}, Swift J0746.3-1608 \citep{Bernardini19}, J183221.56-162724.25 \citep[J1832;][]{beuermann}, and YY Dra itself \citep{Breus17}, though as we noted in Sec.~\ref{sec:2Dpower}, the previously observed YY~Dra low states do not show evidence of a cessation of accretion. Additionally, \citet{covington} have identified low states in V1223~Sgr, V1025~Cen, V515 And, and RX~J2133.7+5107, and they conclude that low states in IPs are more common than previously realized.

Detailed observations of these low states has revealed significant changes in the accretion geometry. FO~Aqr is the best example of this, as optical \citep{Littlefield16, Littlefield20}, X-ray \citep{Kennedy17}, and spectroscopic \citep{Kennedy20} studies have all agreed that the system's mode of accretion changes at reduced mass-transfer rates. This is also true of DW~Cnc and V515~And \citep{covington}. With just one exception, these studies have not detected a total interruption of accretion during low states. 

Of the other IPs that have been observed in low states, only in J1832 has the accretion rate been observed to decline to negligible levels. Similar to YY~Dra, J1832 shows no variability related to the WD's rotation during its deep low states \citep{beuermann}, but J1832 has the added advantage of being a deeply eclipsing system. \citet{beuermann} showed that when J1832 enters a deep low state, its optical eclipse depth becomes imperceptibly shallow, suggesting that there is no remaining accretion luminosity and hence no accretion. Due to gaps in ground-based observational coverage of J1832, it is unclear how long J1832's deep low states last or how long it takes for the system to fall into one. In contrast, the \TESS\ light curve of YY~Dra firmly establishes the duration of the stoppage of accretion (28~h) as well as the length of the abrupt transitions into and out of that episode of negligible accretion (10~h and 2~h, respectively).

Interestingly, J1832 appears to be a persistently diskless IP \citep{beuermann}, and we have argued that YY~Dra's power spectrum suggests that it too was diskless when its accretion ceased. This might be an important clue as to why the accretion rates of other IPs have not been observed to fall to negligible levels during their low states. A Keplerian disk is depleted at its viscous timescale, which can be days or even weeks, but in a diskless geometry, the lifetime of the accretion flow is much shorter (of order the free fall time between the system's L1 point and the WD), so a cessation of mass transfer by the secondary would quickly result in a corresponding stoppage of accretion onto the WD.

More broadly, when looking at low states in other IPs, we see that they tend to last substantially longer (anywhere from months to years) than YY~Dra's 28~h deep low state. The low states in FO Aqr lasted 5-6 months in 2016, 4-5 months in 2017, and only 2-3 months in 2018 \citep{Littlefield20}. Additionally, V1025 Cen \citep{covington, littlefield22} declined 2.5 magnitudes over the course of 4 years and remained in that low state for a little under 2 years. DW~Cnc has significant gaps in the data, but it has shown a nearly 3~year low state with a depth of 2.3 mag \citep{covington}. The lengths of these low states are substantially longer than those of YY Dra, underscoring the remarkable brevity of YY~Dra's deep low state.

Finally, \citet{scaringi17, scaringi21} examined low states of MV~Lyr and TW~Pic observed by \textit{Kepler} and \TESS, respectively. Although both systems are now suspected to be extremely low-field IPs, neither had been known to be magnetic. \citet{scaringi17, scaringi21} identified magnetically gated accretion episodes via the \citet{st93} instability, resulting from a magnetic field strength so low that it produces observable effects (magnetically gated bursts) only during epochs of low accretion. In both MV~Lyr and TW~Pic, the low states tended to last for a number of days or even weeks, although one low state in TW~Pic lasted for just over $\sim1$~d. However, unlike the deep low state of YY~Dra, TW~Pic continued to accrete during this interval.

In conclusion, there is no reported precedent for a $\sim$day-long cessation in accretion in a confirmed IP. During a future episode of near-zero accretion, it should be possible to use optical spectra to search for Zeeman-split photospheric lines from the exposed WD in order to directly measure its magnetic-field strength. 

\subsection{Nature of the Low State}

\citet{livio94} and \citet{hessman} have theorized that star spots can pass over the inner hemisphere of a star and shrink the height of the companion star's atmosphere, causing it to temporarily underfill its Roche lobe. Large star spots will distort the light curve of the companion's ellipsoidal variations. When star spots are present, the shape of the ellipsoidal light curve would be distorted with a dip at certain orbital phases. If the star spots disappear, the dip should go away as well. 

To search for evidence of such a change, we split the deep low state of the \TESS\ light curve into two halves and phased both according to Equation~\ref{ephemeris}. However, there are no statistically significant changes between the first and second halves of the deep low state. This result is perhaps unsurprising, given that each half of the deep low state encompasses only $\sim$3.5 binary orbits, which is too short a baseline to significantly boost the signal-to-noise ratio through phase-averaging the data. However, more precise measurements during a future cessation of accretion might be able to test more closely a change in the ellipsoidal light curve. A major benefit of performing such an analysis in a non-accreting IP (as opposed to a non-accreting polar) is that in a polar, the equality of \spin\ and \orbit\ makes it difficult to disentangle the secondary's ellipsoidal variations from other sources of variability at the orbital frequency, such as accretion or thermal emission from hot spots on the WD surface \citep[e.g., as observed for the low-state polar LSQ1725-64;][]{fuchs}.

Although the \TESS\ data therefore cannot test the starspot model, they do offer a helpful clue as to when mass transfer from the donor (as distinguished from the instantaneous accretion onto the WD) stopped. In a disk-fed geometry, a disk will dissipate at its viscous timescale, which can be several weeks (or longer) at low mass-transfer rates. In such a system, a disk can therefore be present even if it is not being replenished with mass from the donor star, and the only observable sign that the secondary is not transferring mass would be the disappearance of the hotspot where the accretion stream impacts the rim of the disk. Had YY~Dra been accreting from a disk during the \TESS\ observation, we would have expected to detect the hotspot as a contribution to \orbit\ and perhaps its harmonics in the power spectrum, and its disappearance would have been observable in the 2D power spectrum. No such changes are observed. However, in a stream-fed geometry, there is only a short lag between a cessation of mass loss by the companion and the resulting interruption of accretion on to the WD. 

Finally, it is worth noting that the flare at the end of the deep low state bears at least some resemblance to the behavior of the apparently non-magnetic VY~Scl system KR~Aur during its deep low states, when as in YY~Dra ellipsoidal variability dominates the light curve \citep{kr_aur}. However, in KR~Aur, the flaring episodes appear to be short, lasting for tens of minutes as opposed to 0.7~d observed in YY~Dra. Moreover, one of the longer low-state flares in  KR~Aur showed evidence of a quasi-periodicity, while the flare in YY~Dra showed no significant periodicity other than the underlying ellipsoidal variations. 

\section{Conclusion}

During its \TESS\ observation, YY Dra had a unique low state during which accretion turned off altogether for 28~h, leaving behind only ellipsoidal variability in the light curve. Our analysis suggests that both before and after the low state, accretion in YY~Dra was stream-fed, similar to the only other IP that has been detected in a state of negligible accretion. Long-term survey photometry reveals that YY~Dra has experienced episodes of negligible accretion relatively frequently during recent years, which raises the enticing prospect of directly measuring the field strength of the WD via the Zeeman effect during a future low state. 


\acknowledgements
We thank the anonymous referee for an expeditious and helpful report.

PS and CL acknowledge support from NSF grant AST-1514737.

M.R.K. acknowledges funding from the Irish Research Council in the form of a Government of Ireland Postdoctoral Fellowship (GOIPD/2021/670: Invisible Monsters).
\software{astropy \citep{astropy:2013,Astropy2018}, lightkurve \citep{Lightkurve20}, matplotlib \citep{Hunter07}}

\appendix

As noted in the introduction, the IP studied in this paper is confusingly referred to as both YY~Dra and DO~Dra in the literature. Several papers, including \citet{patterson87} and \citet{virnina}, discuss the reasons for this ambiguity. In brief, there is no doubt that ``DO~Dra'' refers to the IP, but ``YY~Dra'' was originally classified as a detached eclipsing binary by \citet{tsesevich}. However, no such object has been identified at or near the nominal position of YY~Dra \citep{patterson87, virnina}. Many previous studies have assumed that YY~Dra was misclassified, that there was a 53~arcsec error in its reported position, and that it is the same object as DO~Dra, as \citet{patterson87} forcefully advocated. However, some studies \citep[e.g.,][]{andronov08} have disagreed, contending that YY~Dra is an eclipsing binary whose position was inaccurately recorded. 

Recent advances in survey photometry afford us a new opportunity to resolve this mystery. Our hypothesis is simple: if YY~Dra were a ``lost'' eclipsing binary star with the properties described by \citet{tsesevich}, it would have been serendipitously recovered by stellar variability surveys, such as the Catalina Surveys Periodic Variable Star Catalog \citep{drake14} and the ASAS-SN Catalog of Variable Stars \citep{jayasinghe18, jayasinghe19}.\footnote{While this paper was under review, Koji Mukai released the 2021 version of his IP catalog on 2021 December 30, and its entry for YY~Dra includes a new description of the naming controversy, available at \url{https://asd.gsfc.nasa.gov/Koji.Mukai/iphome/systems/yydra.html}. Mukai independently used a similar methodology as us to reach the same conclusion: that the ASAS-SN Catalog of Variable Stars contains no stars with orbital periods similar to that reported by \citet{tsesevich}.} In particular, YY~Dra's eclipses were reported to be $\gtrsim1.6$~mag, which would make them deeper than the eclipses of 99\% of the 95,876 Algol-type binaries contained in the VSX catalog at the time of writing;\footnote{For simplicity, we limited our query to stars that are classified solely as a detached eclipsing binary and excluded systems that have multiple types of variability.} such a great eclipse depth, combined with the system's out-of-eclipse brightness (mag 12.9) and orbital period (4.21123~d), should make it readily detectable in existing surveys. We therefore queried the Catalina, ASAS-SN, and VSX catalogs to search for detached eclipsing binaries whose properties agree with those reported by \citet{tsesevich} for YY~Dra. Our search turned up no plausible matches, but we will briefly outline the three scenarios that we considered.

Initially, with no positional constraints, we searched unsuccessfully for eclipsing binaries that match YY~Dra's reported parameters by requiring the peak magnitude to be brighter than $V=14$, the eclipse depth to exceed 1~mag, and the period to be within $\pm0.02$~d of the period reported for YY~Dra. We deliberately selected a loose constraint on the eclipse depth because the ASAS-SN survey has low angular resolution, and the resulting blending would dilute the observed eclipse depth.

We also considered the possibility that \citet{tsesevich} might have underestimated YY~Dra's orbital period by a factor of 2, which could occur if its primary and secondary eclipses were separated by a half-orbit and had similar depths. With this modified constraint,  we found one eclipsing system that agreed reasonably well with YY~Dra's brightness and eclipse depth, but its declination of $-53^{\circ}$ makes it unobservable from Russia, where YY~Dra was discovered.

Finally, it is plausible that the \citet{tsesevich} period was fundamentally erroneous, so we removed the period-based constraint on our search. We found two eclipsing binaries with eclipse depths in excess of 1~mag within an arbitrary $15^{\circ}$ radius of YY~Dra's nominal position, but the out-of-eclipse brightness for each system is incompatible with what was reported for YY~Dra. 

We cannot rule out the possibility that YY~Dra really is an eclipsing binary whose position, period, and range of variation were all massively erroneous. This explanation, however, would require an unlikely confluence of errors and would probably be untestable, in that it would leave the search for YY~Dra almost completely unconstrained.

The lack of an obvious counterpart for YY~Dra in the detached eclipsing binaries in the Catalina, ASAS-SN, and VSX catalogs strongly supports the conclusion of \citet{patterson87} that YY~Dra and DO~Dra are the same system.

\bibliography{Bibliography.bib}

\end{document}